\documentclass[showpacs,preprintnumbers,amsmath,amssymb,twocolumn,floatfix,nofootinbib]{revtex4}
\usepackage[usenames]{color}
\usepackage{epsfig}
\usepackage{amssymb}
\usepackage{xcolor}

\newcommand{\be}{\begin{equation}}
\newcommand{\ee}{\end{equation}}
\newcommand{\bee}{\begin{eqnarray}}
\newcommand{\eee}{\end{eqnarray}}

\definecolor{grey}{rgb}{0.9,0.9,0.9}
\definecolor{black}{rgb}{0,0,0}

\def \irbaddress{Rudjer Bo\v{s}kovi\'{c} Institute, Bijeni\v{c}ka cesta 54, P.O. Box 180, 10002 Zagreb, Croatia}
\def \untzaddress{University of Tuzla, Faculty of Science, Univerzitetska 4, 35000 Tuzla, Bosnia and Herzegovina}

\begin{document}

\title{A new method for extracting poles from single-channel data based on Laurent expansion of T-matrices with Pietarinen power series 
representing the non-singular part}
\author{Alfred \v{S}varc}
\email{alfred.svarc@irb.hr}
\affiliation{\irbaddress}
\author{Mirza Had\v{z}imehmedovi\'{c}}
\affiliation{\untzaddress}
\author{Hedim Osmanovi\'{c}}
\affiliation{\untzaddress}
\author{Jugoslav Stahov}
\affiliation{\untzaddress}

\date{\today}

\begin{abstract}
  We present a new approach to quantifying pole parameters of single-channel processes based on Laurent expansion of partial wave T-matrices. 
Instead of guessing the analytical form of non-singular part of Laurent expansion as it is usually done, we represent it by the convergent 
series of Pietarinen functions. As the analytic structure of non-singular term is usually very well known (physical cuts with branhcpoints at 
inelastic thresholds, and unphysical cuts in the negative energy plane), we show that we need one Pietarinen series per cut, and the number 
of terms in each Pietarinen series is automatically determined by the quality of the fit. The method is tested on a toy model constructed 
from two known poles, various background terms, and two physical cuts, and shown to be robust and confident up to three Pietarinen series. We 
also apply this method to Zagreb CMB amplitudes for the N(1535) 1/2- resonance, and confirm the full success of the method on realistic data. 
This formalism can also be used for fitting experimental data, and the procedure is very similar as when Breit-Wigner functions are used, but 
with one modification: Laurent expansion with Pietarinen series is replacing the standard Breit-Wigner T-matrix form. 
\end{abstract}

\maketitle

The recent Camogli workshop \cite{Camogli2012} has finally inaugurated the fact that poles, and not Breit-Wigner parameters determine and 
quantify resonance properties, and that they should be used a link between scattering theory and QCD. However, at the same time, the question 
of finding an adequate procedure to extract them from single-channel T-matrices has been opened. Experimentalists are quite familiar with 
fitting the data with Breit-Wigner functions (either with constant parameters and very general backgrounds, or with energy dependent mass and 
width), but are inexperienced when actual complex energy poles have to be used. A simple procedure for pole extraction is still missing. Up 
to now poles have usually been extracted from theoretical single or multi-channel models fitted to the data using either of standard pole 
extraction methods: analytic continuation of the model functions into the complex energy plane \cite{Doering,EBAC,CMB,Zagreb,Bonn}, speed 
plot \cite{Hoehler93}, time delay \cite{Kelkar}, N/D method \cite{ChewMandelstam}, regularization procedure \cite{Ceci2008}, etc. However, 
this required solving standardly elaborated single/coupled-channel models and analyzing the obtained analytic solution which implicitly 
contained both parts: singular and background. Hence, the analytic form of the full solution varied from model to model; pole-background 
separation was not unique, and this introduced uncertainties in pole extraction procedures.  The intention of this paper is to offer simple, 
robust and confident method how to obtain scattering matrix poles, but maximally avoid the need to speculate about type and form of 
background terms. We base our analysis on  Laurent expansion of partial wave T-matrices which uniquely separates singular from finite terms, 
and treat singular and finite terms separately. Our main assumption is that all scattering matrix poles are of the first order.

\begin{table*}[p!]
 \caption{Toy model parameters and fitted parameters. Input parameters are given
in boldface, and results of a fit in normal font. Table is given in GeV units, and $\Gamma _i= - 2 \, W_i$. \\ }
\label{tb1:Toy model parameters} %
\begin{tabular}{||c|c|c|c|cccccccc|ccc|ccc|ccc|c||}
\hline
$C_{1}$  & $C_{2}$  & $B_{1}$  & $B_{2}$  & $r_{1}$  & $g_{1}$  & $M_{1}$  & $\Gamma _1$  & $r_{2}$  & $g_{2}$  & $M_{2}$  & $\Gamma _2$  & 
$\alpha$  & $x_{P}$ & $N_{1}$ & $\beta$& $ x_{Q} $&$ N_{2}$  & $\gamma $ & $ x_{R} $&$ N_{3}$  & $10^{2}\chi_{R}^{2}$\tabularnewline
\hline 
\hline 
\multicolumn{4}{||c|}{Toy-model} & \multicolumn{18}{c||}{}    \tabularnewline
\hline 
\multicolumn{4}{||c|}{} & \textbf{ 0.1}  & \textbf{0.09 }  & \textbf{1.65 }  & \textbf{0.165}  & \textbf{0.09}  & \textbf{0.06 }  & 
\textbf{2.25}  & \textbf{0.2}  & \multicolumn{3}{c|}{} & \multicolumn{3}{c|}{} & \multicolumn{3}{c|}{} & \multicolumn{1}{|c||}{} 
\tabularnewline
\hline 
\hline 
\multicolumn{4}{||c|}{Fitted results} & \multicolumn{18}{c||}{}    \tabularnewline
\hline 
\hline 
\multicolumn{4}{||c|}{Strategy \textbf{a.}} & \multicolumn{18}{c||}{}    \tabularnewline
\hline 
1  & 0  & 0  & 0  & 0.100  & 0.089  & 1.649  & 0.165  & 0.090  & 0.060  & 2.249  & 0.200  & 2.48  & 0.97 & 5 & && &  &&& $0.03$ 
\tabularnewline
\hline 
0  & 1  & 0  & 0  & 0.099  & 0.090  & 1.650  & 0.165  & 0.090  & 0.060  & 2.249  & 0.199  & 3.97  & 3.97 & 5 &  &&&  &&& $0.01$ 
\tabularnewline
\hline 
0  & 0  & 1  & 1  & 0.098  & 0.091  & 1.650  & 0.165  & 0.090  & 0.060  & 2.250  & 0.200  & 1.19 & -14.94 & 7 &  & && &&& $0.2$ 
\tabularnewline
\hline 
0  & 0  & -1  & -1  & 0.099  & 0.089  & 1.649  & 0.1649  & 0.089  & 0.059  & 2.249  & 0.199  & 0.99 & -9.63 & 7 &  & && &&& 
$0.01$\tabularnewline
\hline 
\hline 
1  & 0  & 1  & 1  & 0.103  & 0.100  & 1.653  & 0.171  & 0.101  & 0.067  & 2.249 & 0.221  & 0.71 & -0.23 & 11 &  & && &&& $28$ \tabularnewline
\hline 
1  & 0  & 1  & 1  & 0.099 & 0.090  & 1.650  & 0.164  & 0.089  & 0.060  & 2.250  & 0.199  & -2.04 & -17.58 & 5 & 4.27 &0.97 &5  & && & $0.28$ 
\tabularnewline
\hline 
1  & 0  & - 1  & - 1  & 0.097 & 0.087  & 1.651  & 0.161 & 0.090  & 0.060  & 2.250  & 0.201  & 0.90 & -0.39 & 20 &  & && &&&$22.0$ 
\tabularnewline
\hline 
1  & 0  & -1  & -1  & 0.099  & 0.089  & 1.649  & 0.164  & 0.090  & 0.059  & 2.249  & 0.199  & 2.96 & -8.97 & 6 & 1.56 & 0.97 & 6 &&&  & 
$1.00$ \tabularnewline
\hline 
0  & 1  & 1  & 1  & 0.107 & 0.088 & 1.646 & 0.166 & 0.093 & 0.048 & 2.239 & 0.197 & 2.06 & -0.89 & 10 &  &&&  & &&$114.79$\tabularnewline
\hline 
0  & 1  & 1  & 1  & 0.099 & 0.090 & 1.650 & 0.165 & 0.090 & 0.060 & 2.250 & 0.200 & 1.94 & -16.33 & 5 & 6.42 &3.97& 5 & && & 
$0.02$\tabularnewline
\hline 
0  & 1  & -1  & -1  & 0.090 & 0.086 & 1.651 & 0.156 & 0.095 & 0.058 & 2.248 & 0.202 & 0.969 & -0.37 & 12 &  &&&  & &&$238.38$\tabularnewline
\hline 
0  & 1  & -1  & -1  & 0.099 & 0.090 & 1.650 & 0.165 & 0.090 & 0.060 & 2.250 & 0.200 & 0.81 & -7.89 & 8 & 1.24&3.97& 8  & && & 
$0.06$\tabularnewline
\hline 
\hline 
1  & 1  & 1  & 1  & 0.085 & 0.102 & 1.663 & 0.171 & 0.087 & 0.075 & 2.262 & 0.216 & 1.09 & -2.64 & 10 &  & && &&& 328.19\tabularnewline
\hline 
1  & 1  & 1  & 1  & 0.098 & 0.086 & 1.650 & 0.161 & 0.095 & 0.058 & 2.247 & 0.199 & 0.44 & -0.47 & 9 & 1.95 &3.97& 8 & && & 
70.37\tabularnewline
\hline 
1  & 1  & 1  & 1  & 0.099 & 0.090 & 1.650 & 0.164 & 0.089 & 0.061 & 2.251 & 0.200 & 4.19 & -22.99 & 5 & 2.22& 3.98&5 & 1.67& 0.97& 3 & 
0.24\tabularnewline
\hline 
1  & 1  & -1  & -1  & 0.090 & 0.105 & 1.657 & 0.182 & 0.078 & 0.061 & 2.260 & 0.189 & 1.38 & -3.12 & 10 &  & && &&& 467.54\tabularnewline
\hline 
1  & 1  & -1  & -1  & 0.095 & 0.098 & 1.654 & 0.173 & 0.086 & 0.061 & 2.254 & 0.198 & 0.61 & -0.20 & 9 & 25.91& 3.98& 8 & && & 
60.94\tabularnewline
\hline 
1  & 1  & -1  & -1  & 0.100 & 0.090 & 1.650 & 0.165 & 0.090 & 0.060 & 2.250 & 0.200 & 1.85 & -6.25 & 3 & 16.36& 3.97& 3  & 1.32 &0.98& 3 & 
0.72\tabularnewline
\hline 
\hline 
\multicolumn{4}{||c|}{Strategy \textbf{b.}} & \multicolumn{18}{c||}{}    \tabularnewline
\hline 
1  & 0  & 0  & 0  & 0.069 & -0.111 & 1.647 & 0.155 & 0.081 & 0.055 & 2.252 & 0.193 & 0.96 & 0.994 & 18 & && &  &&& 0.83 \tabularnewline
\hline 
0  & 1  & 0  & 0  & -0.101 & -0.084 & 1.649 & 0.165 & -0.088 & -0.061 & 2.249 & 0.200 & 0.90 & 3.94 & 9 &  & && & &&0.33\tabularnewline
\hline 
0  & 0  & 1  & 1  & 0.329 & 0.247 & 1.649 & 0.165 & -0.077 & 0.076 & 2.249 & 0.200 & 1.00 & -0.76 & 8 &  &  &&& &&0.01\tabularnewline
\hline 
0  & 0  & -1  & - 1  & 0.160 & 0.158 & 1.649 & 0.165 & 0.235 & 0.109 & 2.249 & 0.200 & 0.65 & -8.41 & 8 &  &  &&& &&0.01\tabularnewline
\hline \hline
1  & 0  & 1  & 1  & 0.114 & -0.049 & 1.657 & 0.156 & -0.146 & 0.418 & 2.254 & 0.187 & 0.52 & 0.00 & 11 &  &  & &&&&12.3\tabularnewline
\hline 
1  & 0  & 1  & 1  & -0.452 & -0.036 & 1.649 & 0.165 & 0.096 & -0.064 & 2.249 & 0.200 & 1.00 & -5.81 & 5 & 2.06& 0.97& 5 && &  & 
0.01\tabularnewline
\hline 
1 & 0 & -1 & -1 & -0.226 & 0.160 & 1.645 & 0.166 & 0.017 & 0.110 & 2.249 & 0.208 & 0.34 & -0.02 & 20 &  &&& &&& 7.12\tabularnewline
\hline 
1 & 0 & -1 & -1 & 0.116 & -0.010 & 1.650 & 0.164 & 0.226 & -0.202 & 2.249 & 0.199 & 2.14 & -0.18 & 6 & 1.70& 0.98& 6 &  &&& 
0.03\tabularnewline
\hline  
0  & 1  & 1  & 1  & 0.320 & 0.060 & 1.643 & 0.166 & 0.036 & 0.111 & 2.244 & 0.229 & 1.00 & -0.31 & 10 &  & &&&& & 43.59\tabularnewline
\hline 
0  & 1  & 1  & 1  & -0.096 & -0.092 & 1.650 & 0.165 & -0.090 & -0.059 & 2.250 & 0.200 & 1.11 & -3.81 & 5 & 1.60& 3.97& 5&& &  & 
$0.01$\tabularnewline
\hline 
0  & 1  & -1  & -1  & 0.062 & 0.143 & 1.653 & 0.184 & 0.202 & 0.329 & 2.268 & 0.225 & 0.85 & -0.05 & 9 &  & && &&& $102.28$\tabularnewline
\hline 
0  & 1  & -1  & -1  & 0.101 & 0.291 & 1.650 & 0.165 & 0.090 & 0.062 & 2.250 & 0.199 & 4.25 & -66.59 & 6 & 20.03 &3.97& 6  & && & 
$0.01$\tabularnewline
\hline 
\hline 
1  & 1  & 1  & 1  & 0.052 & -0.092 & 1.656 & 0.143 & 0.239 & 0.282 & 2.235 & 0.179 & 1.18 & -0.25 & 12 &  & && &&& 46.02\tabularnewline
\hline 
1  & 1  & 1  & 1  & -0.058 & -0.122 & 1.662 & 0.167 & -0.054 & -0.081 & 2.258 & 0.185 & 0.71 & -0.72 & 6 & 1.38&4.00& 6 & && & 
25.22\tabularnewline
\hline 
1  & 1  & 1  & 1  & -0.318 & 0.258 & 1.648 & 0.165 & 0.073 & -0.104 & 2.247 & 0.207 & 0.67 & -9.20 & 7 & 0.17&3.99& 7 & 0.18 &1.00& 7 & 
1.09\tabularnewline
\hline 
1  & 1  & -1  & -1  & 0.098 & 0.365 & 1.640 & 0.164 & 0.126 & 0.061 & 2.247 & 0.118 & 1.66 & -1.16 & 12 &  & && &&& 25.63\tabularnewline
\hline 
1  & 1  & -1  & -1  & -0.012 & 0.375 & 1.649 & 0.165 & 0.080 & 0.077 & 2.251 & 0.200 & 1.36 & 0.92 & 8 & 2.62& 3.98& 7 & && & 
0.53\tabularnewline
\hline 
1  & 1  & -1  & -1  & -0.088 & 0.069 & 1.65 & 0.164 & -0.015 & 0.370 & 2.249 & 0.201 & 1.11 & -1.62 & 7 & 3.53 & 4.02 &7& 1.56 &0.97 &5 & 
0.23\tabularnewline
\hline \hline
\end{tabular}
\end{table*}

The starting point of our method is Laurent expansion for a function with one pole:
\begin{eqnarray}
\label{eq:Laurent}
T(\omega) &=& \frac{a_{-1}}{\omega_0-\omega}+ \tilde{B}^{L}(\omega);  \, \, \, \, \, \, a_{-1}, \omega_0,\omega \in\mathbb{C}.  
\end{eqnarray}
where $a_{-1}$ and $\omega_0$ are residuum and pole position respectively, and function $\tilde{B}^{L}(\omega)= \sum 
_{n=0}^{\infty}a_n(\omega _0 - \omega)^n$ is regular everywhere in the complex energy plane. However, functions we analyze in reality may and 
do contain other poles for $\omega \neq \omega_ 0$, so if we iterate this procedure we can without loss of generality write down:
\begin{eqnarray}
\label{eq:Laurent}
T(\omega) &=& \sum _{i=1}^{k} \frac{a_{-1}^{(i)}}{\omega _i -\omega}+B^{L}(\omega);    \, \, \, \, a_{-1}^{(i)}, \omega _i, \omega \in  
\mathbb{C}.
\end{eqnarray}
where $k$ is number of poles. $a_{-1}^{(i)}$ and $\omega_ i$ are residua and pole positions for i-th pole respectively, and $B^{L}(\omega )$ 
is a function regular in all $\omega  \neq \omega _i$.

This approach has been already investigated, but the freedom in choosing the exact analytic form of the background contribution $B^{L}(\omega 
)$ has been introducing severe ambiguities.
\newpage
The novelty of our approach is that we propose to avoid discussing the arbitrariness in all possible choices for the background function 
$B^{L}(\omega )$ by replacing it with Pietarinen expansions
 in power series using a complete set of functions with well known analytic properties. 
\\ \\ \noindent 
\emph{ Pietarinen series} \\ \indent
If $F(\omega)$ is a general, unknown analytic function having a cut starting at $\omega=x_P$, then it can be represented in a power series of 
Pietarinen functions in the following way:
\begin{eqnarray}
\label{eq:Pietarinen}
F(\omega ) &=& \sum_{n=0}^{N}c_n\, Z(\omega )^n, \, \, \, \, \, \, \, \, \, \, \omega  \in  \mathbb{C}   \nonumber \\
Z(\omega )&=& \frac{\alpha-\sqrt{x_P-\omega }}{\alpha+\sqrt{x_P-\omega }}, \, \, \, \, \, c_n, x_P, \alpha \in  \mathbb{R},
\end{eqnarray} 
with the $\alpha$ and $c_n$ being tuning parameter and coefficients of Pietarinen function $Z(\omega)$ respectively.  
\\ \\
The Pietarinen series have been proposed and introduced by Ciulli \cite{Ciulli} and Pietarinen \cite{Pietarinen}, and have been with great 
success used in Karlsruhe-Helsinki partial wave analysis \cite{Hoehler84} when invariant amplitudes have been expanded in as many as 50 
terms. The essence of the approach is the fact that a set $(Z(\omega )^n, \, n=1, \, \infty)$ forms a complete set of functions defined on 
the unit circle in the complex energy plane having branch cut starting at $\omega= x_P$, but the analytic form of the function is at the 
beginning yet undefined. The final form of the analytic function $F(\omega)$ is obtained by introducing the rapidly convergent power series 
with real coefficients, and the degree of the expansion is automatically determined in fitting the input data. 
\\ \\
\noindent
\emph{\mbox{The application of Pietarinen series to scattering theory}\vspace{0.1cm}}
\\   \indent
The analytic structure of each partial wave is a well known fact. Each partial wave contains poles which parameterize resonant contributions, 
it has cuts in the physical region starting at thresholds of elastic and all possible inelastic channels, and finally there are
 t-channel, u-channel and nucleon exchange contributions quantified with corresponding negative energy cuts. 
However, explicit analytic form of each cut contribution is not known. Instead of guessing the exact analytic form of all of them, we propose 
to use one Pietarinen series to represent each cut, and the number of terms in Pietarinen series will be determined by the quality of fit to 
the input data. So, in principle we have one Pietarinen series per cut, the branch-points $x_P, x_Q ...$ are known from physics, and 
coefficients are determined by fitting the input data coming from real physical process.  In practice, we have too many cuts (especially in 
the negative energy range), so we reduce their number by dividing them in two categories: all negative energy cuts are approximated with only 
one, effective negative energy cut represented with one Pietarinen (we usually denote its branchpoint as $x_P$), and each physical cut is 
represented with its own Pietarinen series with branch-points determined by the physics of the process ($x_Q,x_R...$).

So, the set of equations which define Laurent expansion+Pietarinen series method (L+P method) is:
\begin{eqnarray}
\label{eq:Laurent-Pietarinen}
T(\omega ) &=& \sum _{i=1}^{k} \frac{a_{-1}^{(i)}}{\omega _i-\omega}+ B^{L}(\omega) \nonumber \\
B^{L}(\omega)&=& \sum _{n=0}^{M}c_n\, Z(\omega )^n  +  \sum _{n=0}^{N}d_n\, W(\omega )^n + \cdots    \nonumber  \\
Z(\omega )&=& \frac{\alpha-\sqrt{x_P-\omega}}{\alpha+\sqrt{x_P - \omega }}; \, \, \, \, \,   W(\omega ) =  \frac{\beta-\sqrt{x_Q-\omega 
}}{\beta+\sqrt{x_Q-\omega }} + \cdots  \nonumber \\
&& a_{-1}^{(i)}, \omega _i, \omega   \in   \mathbb{C} \nonumber \\
&& c_n, x_P, d_n, x_Q, \alpha, \beta ... \in  \mathbb{R}   \nonumber \\ 
&& {\rm and} \, \, \, k, M, N  ... \in  \mathbb{N}.
\end{eqnarray}
 As our input data are on the real axes, the fit is performed only on this dense subset of the complex energy plane. All Pietarinen 
parameters in set of equations (\ref{eq:Laurent-Pietarinen}) are determined by the fit.

Let us observe that the class of  input functions which are convenient to be analyzed with this method is quite wide. 
One may either fit partial wave amplitudes obtained from any of theoretical models, or even experimental data directly. In either case, 
T-matrix is represented by the set of equations (\ref{eq:Laurent-Pietarinen}), minimization function is defined (usually $\chi^2$ type), and 
fitting is done.
\\ \\
\noindent
\emph{Let us summarize our fitting procedure \vspace{0.1cm}}
\\ \indent
First of all let us observe that in the strict spirit of the method, physical Pietarinen branch points $x_Q,x_R,... $ \emph{\textbf{should 
not}} be fitting parameters. Namely, as we have declared that each known cut should be represented by its own Pietarinen series having the 
analytic structure of the very cut, Pietarinen branch points should be fixed to known physical branch points. And this indeed is so in the 
ideal case. However, in realistic case the situation is very similar to the background situation. Namely, we can never include all physical 
cuts in the multi-channel process, so instead of taking them all, we represent them by a smaller subset. So, for physical energy range too, 
Pietarinen branch points $x_Q,x_R,... $ are not constants; we have to relax them and allow them to vary as fitting parameters. Later on in 
this paper we shall demonstrate that when we do it, physical branch points still naturally converge towards branch-points which belong to 
channels which dominate certain partial wave, but do not actually correspond to them exactly. The proximity of the fit results to  exact 
physical branch points describes the "goodness of the fit, namely it tells us how  well certain combination of thresholds indeed is 
approximating certain partial wave. And this, together with the choice of the degree of Pietarinen polynomial represents the model dependence 
of our method. We have, of course,  never claimed that our method is model independent because there is no such thing as model independence. 
However, the method fixes its model dependence naturally, by fitting to the real data. It chooses the simplest function with the given 
analytic properties which fit the data, and increases the complexity of the function only when the actual data require so.  We shall later on 
demonstrate that for 1/2- partial wave ($S_{11}$) the fit chooses $\pi N$ elastic and $\eta $ production branch point, while for 1/2+ partial 
wave ($P_{11}$) instead of choosing  $\eta$ production branch point, it settles close to the $\pi \pi N$ threshold which is dominant for this 
partial wave. Let us observe that we are still limited to the stable two-body channel representation (real branch point), so quasi stable two 
body channels like $\rho N$ and $\sigma N$ are still not included. L+P method can do it by allowing the branch points $x_Q,x_R,... $ to 
become complex numbers, but it simply has not been done yet. 

 We first start with minimal number of poles, the minimal number of Pietarinen functions (we choose only dominant inelastic channels), 
minimal number of fitting parameters $\alpha, \beta, c_n, \, d_n \, ...$, and Pietarinen branch point $x_P,x_Q,... $ being close to actual 
physical branch points. We usually start with $N=3$. The reduced $\chi^2$ is analyzed, and the quality of the fit is visually inspected by 
comparing fitting function with fitted data. If the fit is unsatisfactory (reduced $\chi^2$ is high, or fit visually does not reproduce 
fitted data), the number of Pietarinen parameters $c_n, \, d_n \, ...$ is increased by one. The fit is repeated, and the quality of the fit 
is re-estimated. This procedure is continued until we have reached the sufficient number of Pietarinen terms so that we are able to reproduce 
the fitted data. If the quality of the fit is still unsatisfactory, we first increase the number of poles and repeat the procedure. If no 
improvement is achieved by increasing the number of poles, then we increase the number of Pietarinen functions by one taking the next 
branch-point (next physical threshold), and repeat the procedure until the agreement has been reached.  
\\ \\ \noindent
\centerline{\textbf{Testing the method} }
\\ \\ \noindent
We have tested the method in three ways: \\ \\
I) We have constructed a toy-model function imitating the physical reality as close as possible (known pole parameters, two positive energy 
cuts and various background contributions), and verified how well our method reproduces the input parameters; \\ \\
and to see how our method works in realistic cases \\ \\
 II)  We have used our L+P method to extract pole parameters from single-channel amplitudes of a known, published model (N(1535) 1/2-  
amplitude of Zagreb CMB model \cite{Batinic2010}) and compared the L+P result with pole parameters from the original publication, and \\ \\
III) We have used our L+P method for something what could never be done before, we have extract pole positions from partial wave data of 
ref.~\cite{GWU} without assuming any final functional form for the scattering matrix regular part.
\\ \\ \noindent
\centerline{\emph{I) Fitting the toy model}}
\\ \\ \indent
In principle, we should have defined a toy model input data $T^{ty}(\omega _j)$ by defining a toy model function and by  normally 
distributing its values in order  to simulate the statistical nature of real measured data. However, as the main goal of this paper is to 
establish the validity of the approach, we have restricted our analysis to infinitely precise data by using non-distributed toy-function 
values, and using the statistical weight ${\rm w}_j$ of 5 \%.  This enables us to test the details of our theoretical assumptions, but gives 
unrealistically low $\chi^ 2$ values. Testing the capacity and limitations of the L+P method with realistic data for the toy model is 
deferred to another publication. 
\\ \\ \indent
A toy model function is constructed by assuming a typical analytic structure of a partial wave: it is constructed as a sum of two poles, two 
physical cuts and several non-resonant background contributions.  The function representing physical cuts  is constructed from a function 
$\Re e(x,a) = \sqrt{x^2-4 a x}/2x$ having a cut starting from $x_0= 4 a$  on the real axes\footnote{The type of the function used for 
physical cuts comes from the phase space factor for two body reactions $\phi(s)=\sqrt{\Lambda(s)}/2s$, with $\Lambda(s)=s^2 - 2 s 
(M^2+m_\pi^2)+(M^2-m_\pi^2)^2$, and taking $m_\pi=M$.}, and the analyticity is imposed through the once subtracted dispersion relation 
\mbox{$\Phi(x,a)=\frac{x-x_0}{\pi}\int_{x_0}^{\infty}\frac{\Re e(x',a)}{(x'-x)(x'-x_0)} \, dx'$}.
However, to simplify the demonstration of usability of L+P method, we replace all negative energy cuts with two poles deep in unphysical 
region. In spite of looking rather restrictive, such an approximation is fairly justified. Namely, we know that each cut can be represented 
by the infinite sum of poles, and as negative cut is indeed very far from the region of interest, replacing it with only two out of infinite 
number of poles is a good approximation (see Cutkosky CMB approach \cite{CMB}). 
\\ \noindent
So our toy -model function is given as: 
\begin{eqnarray}
T^{ty}(\omega )  &=& \sum _{i=1}^{ 2} \frac{r^{ty}_{i} + \imath \, g^{ty}_{i}}{ M^{ty}_{i}-\omega  - \imath \, W^{ty}_{i}} +  \\ 
                 &+ & C_1 \, \Phi(\omega,0.25)+ C_2 \,  \Phi(\omega,1.)+ B^{ty}(\omega ),  \nonumber \\ 
 \Phi(\omega,a)&=& \frac{\sqrt{\omega (-4 a + \omega)}}{2 \pi \omega} \ln  \frac{{2 a - \omega - \sqrt{\omega(-4 a + \omega)}}}{2a} \nonumber 
\\
B^{ty}(\omega ) & = & B_1 \,  \frac{10.}{ -10. -\omega - \imath \,  5.}+ B_2 \,  \frac{10.}{ -6. -\omega - \imath \, 4.}, \nonumber 
\end{eqnarray}
where 
\begin{eqnarray}
 r^{ty}_{i},g^{ty}_{i},M^{ty}_{i},W^{ty}_{i} \in \mathbb{R}. & \, \, \, \, \, \, \, \,  &   \nonumber 
\end{eqnarray}
Toy model parameters for all test cases are chosen to resemble physical reality as much as possible, and are given in Table \ref{tb1:Toy 
model parameters} with bold face characters. 
\begin{figure}[!t]
\includegraphics[width=9cm]{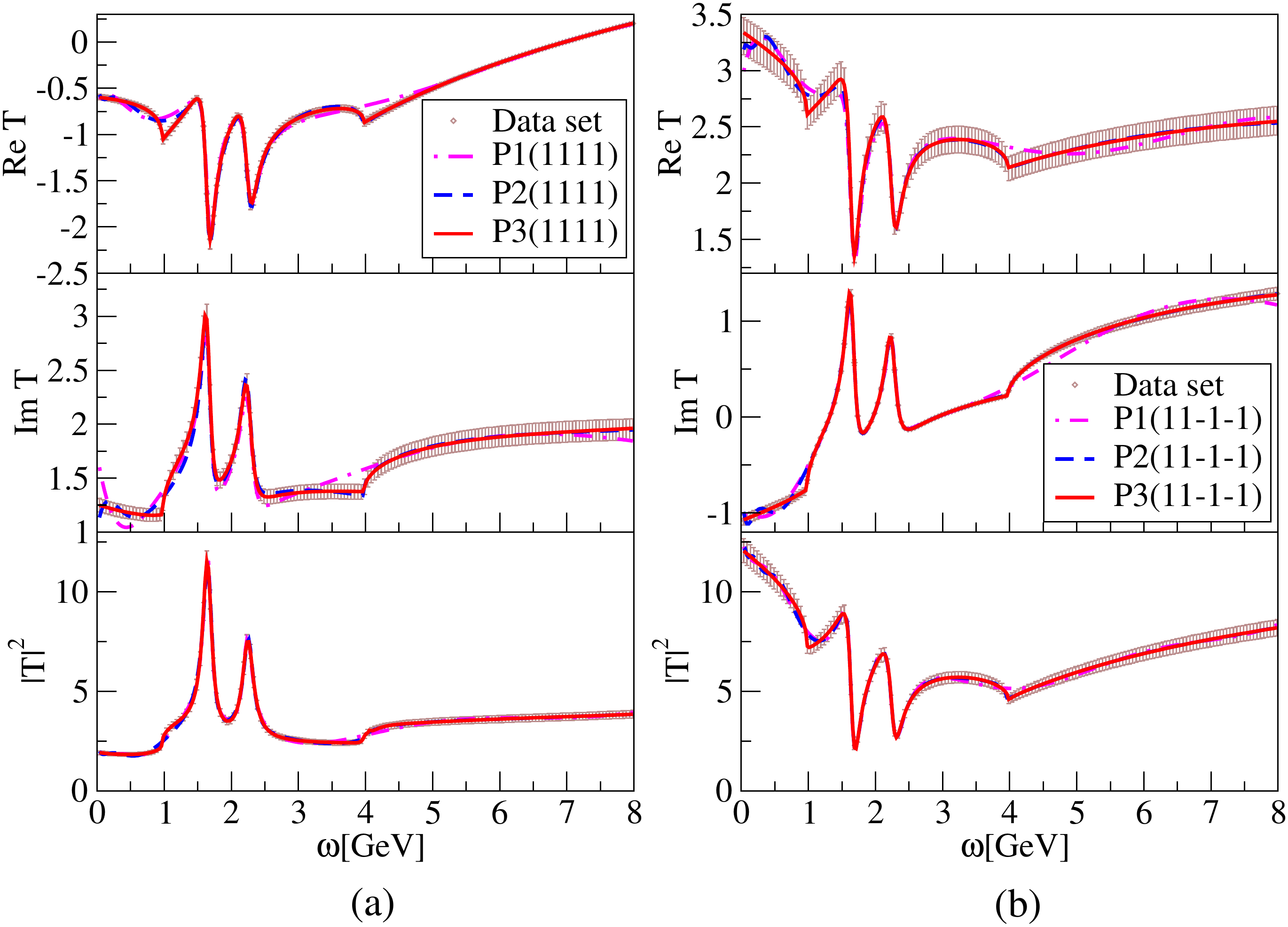}  
\caption{\small \textbf{Toy model function.} In figures a) and b) we give toy-model data for the function with two poles and two cuts for two 
choices of background parameters, for  ($B_1$ = 1, $B_2$ = 1) and  ($B_1$ = -1, $B_2$ = -1) respectively. Dashed-dotted, dashed and full 
lines P1, P2 and P3 give the quality of the fit for solutions with one, two and three Pietarinen expansions respectively. }
\label{Fig1} 
\end{figure}
\\ \\ \noindent
\emph{We have applied two fitting strategies:} \\ \\ \textbf{a)} fitting both, real and imaginary part of the toy-model data (imitating the 
physical situation when a complete experiment is performed, and the full T-matrix is known); and \\ \\ \textbf{b)} fitting only absolute 
value of the toy-model function (imitating physical situation when only incomplete data set is available, and these are usually differential 
cross section data).  \\ \\ \noindent
Minimization function for case a):
\begin{eqnarray}
\chi^2 & =& \chi_f + \lambda \, \chi_{Pen};  \nonumber \\
\chi_f &=&  \sum _{j=1}^{N_{pts}} \mid T^{ty}(\omega _j) - T(\omega _j) \mid ^2  /{\rm w}_{j}^2.  
  \end{eqnarray}
  
\noindent 
Minimization function for case b):
\begin{eqnarray}
\chi^2 & =& \chi_f + \lambda \, \chi_{Pen};   \nonumber \\
 \chi_f &=&  \sum _{j=1}^{N_{pts}}\left| \mid T^{ty}(\omega _j) \mid^2- \mid T(\omega _j) \mid ^2 \right| /{\rm w}_{j}^2.  
 \end{eqnarray}
In both cases ${\rm w}_{j}$ is corresponding statistical weight and $\chi_{Pen} =  \sum _{k=1}^{N} c_k^2 \, k^3$ is Pietarinen series penalty 
function \cite{Pietarinen} which guarantees that the number of power-series terms is minimal. Coefficient $\lambda$ is determined in a fit, 
and serves to match the size of $\chi_f$ versus $\chi_{Pen}$ contributions. 

In situation a) we expect a unique solution, and in situation b) we expect a full set of solutions because the relative phases of the fitted 
functions are still undetermined. 

Let us introduce a notation [a,b,c,d] meaning by definition: ($C_1$= a, $C_2$= b, $B_1$= c, $B_2$= d). 

 We have tested the validity of the model for two backgrounds labeled A=[a,b,1,1] and B=[a,b,-1,-1] which strongly contribute to the 
toy-function strength in the observed resonance range ($ 1 \leq \omega \leq 3$ GeV), and produce drastically different form of the 
toy-function absolute value [see Fig.~(\ref{Fig1})]. We have decided to characterize the type of the background according to the absolute 
value shape (what basically corresponds to the differential cross section): while background A produces typical "two peak" resonance 
structure, background B produces a very atypical behavior in the  first-second resonance region. If these numbers were differential cross 
sections extracted from experiment, one could not easily say whether they in case of background B represent a resonance, or some other 
interference effect. Therefore, we believe that these two backgrounds are the worst case scenario for the L+P method to separate resonance 
and background contributions. \\ \\
\noindent
\underline{Results for fitting strategy a)} \\ \\ \indent 
With toy model data we tested the functionality of the model for various combinations of toy-model ingredients. The aim is to verify basic 
concepts of L+P method for the ideal data set.  All results of the toy-model L+P fit are shown in Table \ref{tb1:Toy model parameters}.  Toy 
model function with all ingredients included, and for two different background contributions ($B_1$ =  1, $B_2$ =  1) and ($B_1$ =  -1, $B_2$ 
= -1) is depicted in Figs.~(\ref{Fig1} a. and b.).
\\
 \noindent
{\em  Two poles, no background one cut} \\ 
We fit the toy model data generated by the toy function with [1,0,0,0] and [0,1,0,0]. We establish that one Pietarinen expansion in L+P 
formalism is sufficient to reproduce the input data (see low  $\chi^2$ value). However, Pietarinen branch point $x_P$ differs for both 
solutions (1 and 4) indicating that the cut structure is reproduced. Both, residua and pole positions are perfectly reproduced. \\ \noindent
 {\em  Two poles, background, no cuts} \\ 
We fit the toy model data generated by the toy function with [0,0,1,1] and [0,0,-1,-1]. We establish that one Pietarinen expansion in L+P 
formalism with cut at $x_P \ll$ 0 is sufficient to reproduce the input data (see low $\chi^2$ value).  Pietarinen branch point $x_P$ differs 
for both solutions indicating that the different backgrounds are fitted. Both, residua and pole positions are perfectly reproduced.
 \\ \noindent
{\em  Two poles first  cut, first background} \\ 
We fit the toy model data generated by the toy function with [1,0,1,1] (background A). We see that for one Pietarinen expansion reduced  
$\chi^2$ value is high, and the Pietarinen branch point is negative. To improve the fit we had to introduce second Pietarinen series. The fit 
is improved, and the fitting has converged.  So, in spite of the fact that we have represented the negative cut as a sum of unphysical poles, 
we still see it as a cut, so we need two Pietarinen expansions to take both cuts into account. While in case of one Pietarinen the 
branchpoint is negative, in case of two Pietarinens one branch point is negative, and second one converges towards 1 what is the toy-function 
branch point.  Both, residua and pole positions are perfectly reproduced. \\
\noindent
{\em  Two poles first cut, second background} \\ 
We fit the toy model data generated by the toy function with [1,0,-1,-1] (background B). The situation of as for first background [1,0,1,1] 
is reproduced. We again need two Pietarinen series. Overall conclusion is that regardless of the type of background, L+P formalism works. 
Both, residua and pole positions are perfectly reproduced.
\\ \noindent
{\em  Two poles second cut, both backgrounds} \\
We fit the toy model data generated by the toy function with [0,1,1,1] and [0,1,-1,-1].
Situation is exactly reproduced as for first cut, but with one difference: Pietarinen branchpoint $x_P$ for two Pietarinen expansions 
converges towards $x_P$ = 4 what is exactly second toy-model cut. Both, residua and pole positions are perfectly reproduced.
\\ \noindent
{\em  Two poles two cuts, first background} \\ 
We fit the toy model data generated by the toy function[1,1,1,1] first with only one Pietarinen expansion, and afterwards with two.  In both 
cases reduced $\chi^2$ value is, as expected, high. The value of $\chi^2$ is better for two Pietarinens, but the result is still 
unsatisfactory. Adding third Pietarinen fixes the problem. The fit has converged, and Pietarinen cuts are consistent with expectations:  
first one is negative, and second and third are close to 1 and 4, namely close to toy-model branch-points.  Both, residua and pole positions 
are perfectly reproduced.
\\ \\ \noindent
{\em  Two poles two cuts, second background} \\
Conclusions are identical as for the first background. Consequently, the L+P formalism is invariant with respect to the relative size and 
phase between background and physical contributions. \\ \\ \noindent
\underline{Results for fitting strategy b)} \\ \\  \noindent 
All conclusions as for fitting strategy a) are reproduced with one major exception: \\ we only reproduce pole positions, residua are quite 
arbitrary. As a matter of fact, when fitting strategy b) was used, we have been able to produce a whole class of solutions with almost 
identical $\chi^2$ and very different residua, exactly what was to be expected. \\ \\
\begin{center}
\emph{II) Fitting $\pi N$ elastic N(1535) 1/2-} 
\end{center}
To illustrate how usable L+P method is in reality, we have decided to fit one realistic set of input data. 
We have taken $\pi N$ elastic N(1535) 1/2-  partial wave from Zagreb CMB model, and fitted it with L+P method. We have chosen fitting 
strategy a) and fitted both, real and imaginary part. We have to mention that we did not have much use of the Penalty function in fitting the 
toy-model, but for realistic data from Zagreb CMB using the penalty function was unavoidable. Results are shown in Table \ref{tb2:S11} and 
Fig. (\ref{Fig2}).
\begin{table*}[t!]
 \caption{Parameters of the fit for the $\pi N$ elastic N(1535) 1/2- partial
wave from Zagreb CMB model. Table is given in MeV units, and $\Gamma _i= - 2 \, W_i$.\\}
\label{tb2:S11} %
\begin{tabular}{|c|cc|cc|cc|c|c|c|c|}
\hline 
 & M$_{1}$  & $\Gamma_{1}$  & M$_{2}$  & $\Gamma_{2}$  & M$_{3}$  & $\Gamma_{3}$  & $x_{P}$, N$_{1}$  & $x_{Q}$, N$_{2}$  & $x_{R}$, N$_{3}$  
& $10^{2}\chi_{R}^{2}$ \tabularnewline
\hline 
\hline 
Zagreb CMB  & \textbf{1521(14) }  & \textbf{ 190(28)}  & \textbf{1646(8)}  & \textbf{204(17) }  & \textbf{1790(26) }  & \textbf{420(45)}  & 
\multicolumn{1}{c}{} & \multicolumn{1}{c}{} &  & \tabularnewline
\hline 
Fit: $_{{\rm 3\, Pietarinens}}^{{\rm 2\, poles}}$  & 1525  & 120  & 1653  & 189  & -  & -  & -696, 15  & 1058, 15 & 1484, 15 & 
2.53\tabularnewline
\hline 
Fit: $_{{\rm 3\, Pietarinens}}^{{\rm 3\, poles}}$  & 1529 & 146  & 1647  & 192  & 1801 & 2321  & 933, 15 & 1057, 15  & 1482, 15   & 1.31 
\tabularnewline
\hline 
\end{tabular}
\end{table*}
\begin{figure}[!t]
\includegraphics[width=8cm]{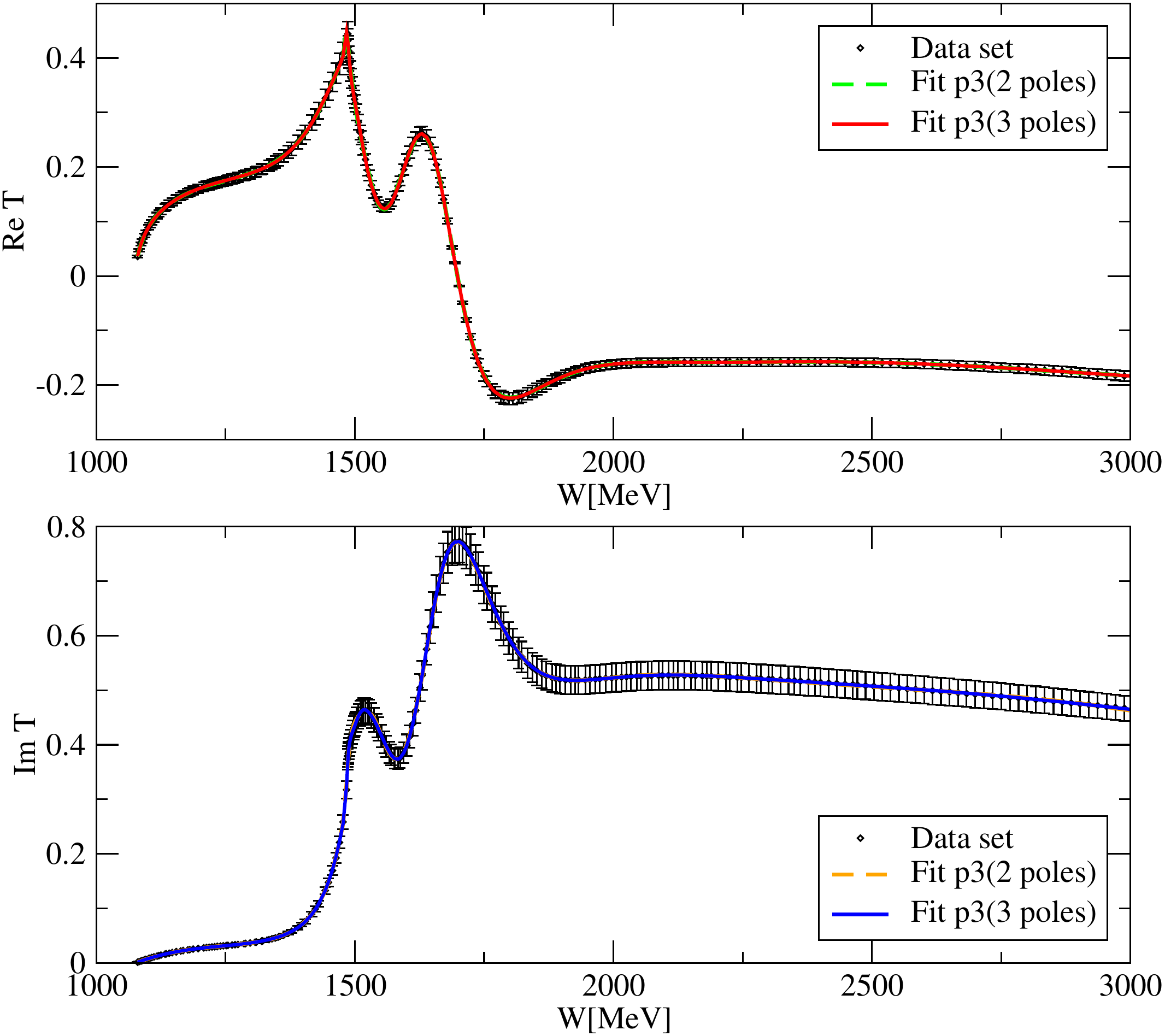}  
\caption{\small \textbf{Zagreb S11.}  Dashed and full lines gives the quality of the fit for two and three pole solutions in both cases with 
three Pietarinen series. }
\label{Fig2} 
\end{figure}

We know that $\pi N$ elastic scattering has at least two branch-points in the physical region ($\pi N$ elastic threshold in Zagreb model at w 
= 1.076 GeV and $\eta$ production In Zagreb model at w = 1.486 GeV), background contribution,  and at least two poles. So we know that our 
L+P solution should have minimally two poles and three Pietarinen series. Therefore, we have started the fitting procedure with two poles, 
and three Pietarinen series, and obtained a reasonable solution. The solution was acceptable visually, and by $\chi ^2$. Two Pietarinen 
branch-points came out very close to physical thresholds, the third Pietarinen branch-point came out far in the negative energy plane 
describing the background contribution from the negative cut, and pole parameters have been quite close to parameters obtained in analytic 
continuation.  We wondered if the fit could be improved by increasing the number of poles by one, so we have tried a three pole-three 
Pietarinen fit.  The reduced $\chi ^2$ was somewhat improved, the pole parameters came somewhat closer to the analytic continuation value, 
but in general the third resonance was uncertain and rather wide. It is interesting to observe that both physical Pietarinen branch-points 
remain where they were (close to physical thresholds), but the third came out much higher. So, L+P method conclusively gives two resonances 
in Zagreb CMB amplitudes, but indicates the existence of the third one which is not well defined. 

We ended up with using three Pietarinen series with one threshold in unphysical range representing the Pietarinen series for background 
contributions, and two thresholds in the physical range. One was at \mbox{$x_Q=1.057$~Gev} (near physical threshold), and second at 
\mbox{$x_R=1.482$~Gev} (near $\eta$ production threshold). Pole positions of Zagreb model are  almost perfectly reproduced. Using L+P method 
we obtained three poles: \mbox{$M_1=1.529 - i \, 0.073$},  \mbox{$M_2=1.647 - i \, 0.096$} and \mbox{$M_3=1.801 - i \, 1.160$ GeV}, what is 
to be compared to Zagreb CMB poles from ref. \cite{Batinic2010}: \mbox{$M_1^{Zg}=1.521 - i \, 0.095$},  \mbox{$M_2^{Zg}=1.646 - i \, 0.102$} 
and \mbox{$M_3^{Zg}=1.790 - i \, 0.210$ GeV}.  First two poles agree almost ideally, the existence of third pole around 1.8 Gev is allowed, 
but single-channel data are insufficient to pin it down more precisely as expected from coupled-channel calculation.

As Zagreb CMB analytic continuation was performed in three channel model, we conclude that two resonances is the best what a single-channel 
method like L+P can give. We also allow for some deviations in quantitative values of two well established resonances, since L+P is, again,  
only single-channel model. A correct recipe would be to repeat the L+P procedure on amplitudes from all channels, and then make an analysis. 
Extending the analysis to other channels might shift pole position somewhat deeper in the complex energy plane, single channel analysis as we 
did it now is, however, "satisfied" with poles being somewhat closer to the real axes. 
\begin{center}
\emph{III) Poles from experiment:} \\
\emph{Fitting GWU-SES partial wave  data}
\end{center}

The novelty of L+P method is that it allows extracting poles from data coming directly from experiment, i.e. to analyze the numbers which are 
obtained using only statistical methods with very little underlying theory. We shall illustrate this feature on GWU single energy (SES), and 
energy dependent (SP06) solutions \cite{GWU}. 

We have extracted pole positions from both, GWU-SES ("experiment") and energy dependent SP06 solution which is obtained from GWU-SES with the 
theoretical analysis based on using the polynomial form of Chew-Mandelstam K-matrix model, and compared results.

Let us point out that GWU group is able to extract poles from their SES only through creating their energy dependent SP06 solution, and 
analytically continuing it into complex energy plane. Our method allows us to fit both, SES and SP06, and compare results. 
\\ \noindent
We came to a very interesting conclusion: \\ \noindent
We can fit both, GWU-SES and SP06, by using  two poles and three Pietarinen series, first threshold being in unphysical range, second near 
physical threshold and third near $\pi \pi N$ threshold (observe that S$_{11}$ was fitted with Pietarinen near $\eta$ production threshold). 
Results of a fit are given in Table~\ref{tab:GWU}. In boldface we give pole positions from original publication. The original publication 
finds Roper poles on two Riemann sheets (on P[221] and on P[121]), and we give them both. However, our L+P method is restricted to the first 
Riemann sheet only. 
\begin{figure}[!t]
\includegraphics[width=8cm]{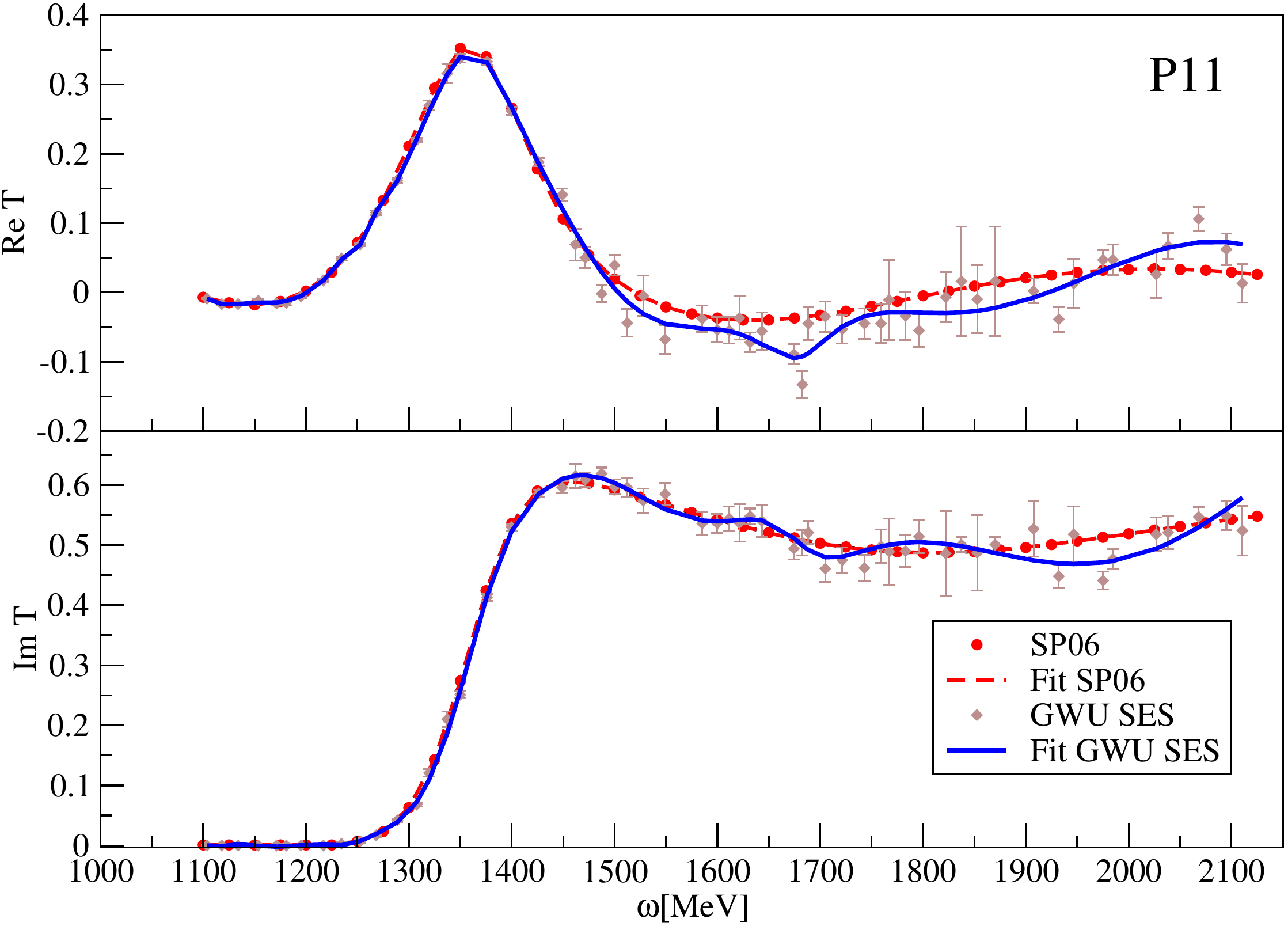}  
\caption{\small GWU SP06 and SES.}
\label{Fig2} 
\end{figure}

\begin{table*}[t!]
 \caption{GWU SES and SP06 from ref. \cite{GWU}. Poles from original publication are given in boldface. Table is given in GeV units, and 
$\Gamma _i= - 2 \, W_i$. \\ }
\label{tab:GWU} %
\begin{tabular}{||c|c|c|c|cccccccc|ccc|ccc|ccc|c||}
\hline
\multicolumn{4}{||c|}{} & $r_{1}$  & $g_{1}$  & $M_{1}$  & $\Gamma _1$  & $r_{2}$  & $g_{2}$  & $M_{2}$  & $\Gamma _2$  & $\alpha$  & $x_{P}$ 
& $N_{1}$ & $\beta$& $ x_{Q} $&$ N_{2}$  & $\gamma $ & $ x_{R} $&$ N_{3}$  & $\chi_{R}^{2}$\tabularnewline
\hline 
\multicolumn{4}{||c|}{GWU SES} & \textbf{ -}  & \textbf{- }  & \textbf{- }  & \textbf{-}  & \textbf{-}  & \textbf{- }  & \textbf{-}  & 
\textbf{-}  & \multicolumn{10}{|c|}{} \tabularnewline
\hline
\multicolumn{4}{||c|}{GWU SP06} & \textbf{-}  & \textbf{- }  & {\boldmath$^{1.388}_{1.358}$}  & \boldmath$^{164}_{162}$  & \textbf{-}  & 
\textbf{-}  & \textbf{-}  & \textbf{-}  & \multicolumn{10}{|c|}{}  \tabularnewline
\hline \hline
 \multicolumn{4}{||c|}{L+P SES} & 0.008 & -0.059 & 1.358 & 0.156 & -0.001 & 0.006 & 1.679 & 0.106 & 1.109 &  -2.409  & 15 & 0.573 & 1.074 & 
15 & 0.677 &  1.243 & 15 & 1.815   \tabularnewline 
\hline
 \multicolumn{4}{||c|}{L+P SP06} & 0.001 & -0.065 & 1.357 & 0.188 & 0.180 & 0.504 & 2.694 & 1.786 & 0.987 & -1.195 & 15 & 1.125 & 1.024 & 15 
& 0.711 &  1.208  & 15 & 0.005   \tabularnewline 
\hline \hline
\end{tabular}
\end{table*}
 From Table~\ref{tab:GWU} we see that for  SP06 the second pole  in our fit is completely undetermined, while for GWU-SES it is definitely 
established, and close to 1.7 GeV.   And that is very similar to what GWU group claims: they claim that they do not need P$_{11}(1710)$ in 
their analysis at all. However, their analysis is based solely on energy dependent analysis.  We agree with them in a sense that  we also do 
not need P$_{11}(1710)$ to fit their energy dependent solution SP06. However, when fitting GWU-SES what they could not do, our fit 
\mbox{\textbf{REQUIRES}}  the presence of the $P_{11}(1710)$ state! Our results are:  \mbox{$M_1^{\rm SP06}=1.358 - i \, 0.094$},  
\mbox{$M_2^{\rm SP06}=2.694 - i \, 0.893$ GeV} for SP06  (second resonance completely undetermined), and \mbox{$M_1^{\rm SES}=1.358 - i \, 
0.078$},  \mbox{$M_2^{\rm SES}=1.679 - i \, 0.053$ GeV} for SES (second resonance firmly established). 

So, our L+P method allows us to explicitly see that their SES contain P$_{11}(1710)$, and their energy dependent analysis smeared it out.  
Results of a fit are given in  Fig.~\ref{Fig2}.

Let us also observe that taking only two Pietarinen series for physical range is  still an approximation. Namely, in both fits given in Table 
\ref{tab:GWU} second and third threshold are close, but not exactly on top of first two physical thresholds. The first  threshold for SES is 
very close to $\pi N$ channel opening (at 1.076 GeV), while the second one is notably above $\pi \pi N$ threshold (at 1.218 GeV). For SP06 
fit thresholds show opposite trend - first threshold is significantly lower, and second threshold is very close to $\pi \pi N$ value. That 
indicates the level of precision of our pole results, and stimulates further improvement of the model. However, even this level of precision 
is sufficient for pursuing the main argument . 
 \\ \\ 
\noindent
\emph{Conclusion} \\ \indent
We propose a new single-channel pole extraction method based on Laurent decomposition and Pietarinen series. 
Instead of guessing the unknown T-matrix functional form as it has been done up to now, we propose to use the Laurent's expansion of 
single-channel partial wave T-matrices to obtain their pole positions. We make use of a fairly well known analytic structure of each partial 
wave, and expand the unknown regular part of the Laurent expansion in power series of Pietarinen functions using one Pietarinen series for 
each known cut. Finally, we  fit  the input data and get the simplest function for the regular Laurent part which has the correct analytic 
structure and together with the known number of poles reproduces the input. The resulting pole parameters are the genuine poles of our 
unknown input function. Even to the surprise of the authors, the fitting procedure in analyzed test cases turns out to be  extremely stable 
and reliable regardless of the number of Pietarinen series, and extracted pole parameters turn out to be fairly confident.  We warn the 
reader that this is a \emph{single-channel method}, and it will recognize only those resonances which strongly couple to the observed 
channel. Other resonances will be only hinted at, and have to be pinned down by analyzing other channels.

\begin{table*}[t!]
\begin{center}
{\huge \textbf{Appendix}}
\end{center}
{Pietarinen expansion coefficients of solutions presented
in Table \ref{tb1:Toy model parameters} \vspace*{0.5cm} (\textbf{Strategy a}).}
\label{tb2:Coefficients Strategy a}
\footnotesize

\begin{tabular}{|c|c|c|c|ccccccccccc|}
\hline \hline
\multicolumn{4}{|c}{Startegy a:} & \multicolumn{11}{c|}{One Pietarinen expansion}\tabularnewline
\hline \hline
$C_{1}$ & $C_{2}$ & $B_{1}$ & $B_{2}$ & $\begin{array}{c}
c_{0}\\
c_{11}
\end{array}$ & $\begin{array}{c}
c_{1}\\
c_{12}
\end{array}$  & $\begin{array}{c}
c_{2}\\
c_{13}
\end{array}$  & $\begin{array}{c}
c_{3}\\
c_{14}
\end{array}$ & $\begin{array}{c}
c_{4}\\
c_{15}
\end{array}$  & $\begin{array}{c}
c_{5}\\
c_{16}
\end{array}$  & $\begin{array}{c}
c_{6}\\
c_{17}
\end{array}$  & $\begin{array}{c}
c_{7}\\
c_{18}
\end{array}$ & $\begin{array}{c}
c_{8}\\
c_{19}
\end{array}$  & $\begin{array}{c}
c_{9}\\
c_{20}
\end{array}$ & $\begin{array}{c}
c_{10}\\
c_{21}
\end{array}$\tabularnewline
\hline \hline
1 & 0  & 0  & 0  & $\begin{array}{c}
1.079\\
-
\end{array}$ & $\begin{array}{c}
-1.080\\
-
\end{array}$ & $\begin{array}{c}
0.155\\
-
\end{array}$ & $\begin{array}{c}
-0.175\\
-
\end{array}$ & $\begin{array}{c}
0.044\\
-
\end{array}$ & $\begin{array}{c}
-0.022\\
-
\end{array}$ & $\begin{array}{c}
-\\
-
\end{array}$ & $\begin{array}{c}
-\\
-
\end{array}$ & $\begin{array}{c}
-\\
-
\end{array}$ & $\begin{array}{c}
-\\
-
\end{array}$ & $\begin{array}{c}
-\\
-
\end{array}$\tabularnewline
\hline 
0  & 1  & 0  & 0  & $\begin{array}{c}
0.947\\
-
\end{array}$ & $\begin{array}{c}
-1.031\\
-
\end{array}$ & $\begin{array}{c}
0.205\\
-
\end{array}$ & $\begin{array}{c}
-0.153\\
-
\end{array}$ & $\begin{array}{c}
0.044\\
-
\end{array}$ & $\begin{array}{c}
-0.012\\
-
\end{array}$ & $\begin{array}{c}
-\\
-
\end{array}$ & $\begin{array}{c}
-\\
-
\end{array}$ & $\begin{array}{c}
-\\
-
\end{array}$ & $\begin{array}{c}
-\\
-
\end{array}$ & $\begin{array}{c}
-\\
-
\end{array}$\tabularnewline
\hline 
0  & 0  & 1  & 1  & $\begin{array}{c}
26.944\\
-
\end{array}$ & $\begin{array}{c}
62.508\\
-
\end{array}$ & $\begin{array}{c}
20.310\\
-
\end{array}$ & $\begin{array}{c}
-51.562\\
-
\end{array}$ & $\begin{array}{c}
-20.914\\
-
\end{array}$ & $\begin{array}{c}
55.819\\
-
\end{array}$ & $\begin{array}{c}
58.087\\
-
\end{array}$ & $\begin{array}{c}
17.230\\
-
\end{array}$ & $\begin{array}{c}
-\\
-
\end{array}$ & $\begin{array}{c}
-\\
-
\end{array}$ & $\begin{array}{c}
-\\
-
\end{array}$\tabularnewline
\hline 
0  & 0  & -1  & -1  & $\begin{array}{c}
-54.437\\
-
\end{array}$ & $\begin{array}{c}
-198.58\\
-
\end{array}$ & $\begin{array}{c}
-266.51\\
-
\end{array}$ & $\begin{array}{c}
-95.229\\
-
\end{array}$ & $\begin{array}{c}
150.39\\
-
\end{array}$ & $\begin{array}{c}
212.01\\
-
\end{array}$ & $\begin{array}{c}
111.12\\
-
\end{array}$ & $\begin{array}{c}
22.976\\
-
\end{array}$ & $\begin{array}{c}
-\\
-
\end{array}$ & $\begin{array}{c}
-\\
-
\end{array}$ & $\begin{array}{c}
-\\
-
\end{array}$\tabularnewline
\hline 
1 & 0  & 1  & 1  & $\begin{array}{c}
-2.199\\
-
\end{array}$ & $\begin{array}{c}
-2.705\\
-
\end{array}$ & $\begin{array}{c}
-2.005\\
-
\end{array}$ & $\begin{array}{c}
-1.117\\
-
\end{array}$ & $\begin{array}{c}
0.402\\
-
\end{array}$ & $\begin{array}{c}
1.497\\
-
\end{array}$ & $\begin{array}{c}
2.562\\
-
\end{array}$ & $\begin{array}{c}
2.329\\
-
\end{array}$ & $\begin{array}{c}
1.827\\
-
\end{array}$ & $\begin{array}{c}
0.918\\
-
\end{array}$ & $\begin{array}{c}
0.424\\
-
\end{array}$\tabularnewline
\hline 
1  & 0  & - 1  & - 1  & $\begin{array}{c}
4.011\\
0.900
\end{array}$ & $\begin{array}{c}
2.215\\
1.289
\end{array}$ & $\begin{array}{c}
3.875\\
-0.409
\end{array}$ & $\begin{array}{c}
2.001\\
0.691
\end{array}$ & $\begin{array}{c}
2.916\\
0.750
\end{array}$ & $\begin{array}{c}
-0.045\\
1.967
\end{array}$ & $\begin{array}{c}
0.723\\
1.710
\end{array}$ & $\begin{array}{c}
-1.118\\
1.584
\end{array}$ & $\begin{array}{c}
1.620\\
0.746
\end{array}$ & $\begin{array}{c}
0.861\\
0.307
\end{array}$ & $\begin{array}{c}
2.834\\
-
\end{array}$\tabularnewline
\hline 
0 & 1  & 1  & 1  & $\begin{array}{c}
-2.798\\
-
\end{array}$ & $\begin{array}{c}
0.641\\
-
\end{array}$ & $\begin{array}{c}
-6.514\\
-
\end{array}$ & $\begin{array}{c}
5.436\\
-
\end{array}$ & $\begin{array}{c}
-11.528\\
-
\end{array}$ & $\begin{array}{c}
8.090\\
-
\end{array}$ & $\begin{array}{c}
-11.12\\
-
\end{array}$ & $\begin{array}{c}
5.779\\
-
\end{array}$ & $\begin{array}{c}
-5.968\\
-
\end{array}$ & $\begin{array}{c}
1.872\\
-
\end{array}$ & $\begin{array}{c}
-1.377\\
-
\end{array}$\tabularnewline
\hline 
0  & 1  & -1  & -1  & $\begin{array}{c}
3.538\\
0.448
\end{array}$ & $\begin{array}{c}
1.523\\
0.479
\end{array}$ & $\begin{array}{c}
2.114\\
-
\end{array}$ & $\begin{array}{c}
0.383\\
-
\end{array}$ & $\begin{array}{c}
1.654\\
-
\end{array}$ & $\begin{array}{c}
-0.325\\
-
\end{array}$ & $\begin{array}{c}
1.1333\\
-
\end{array}$ & $\begin{array}{c}
0.130\\
-
\end{array}$ & $\begin{array}{c}
1.499\\
-
\end{array}$ & $\begin{array}{c}
0.773\\
-
\end{array}$ & $\begin{array}{c}
1.345\\
-
\end{array}$\tabularnewline
\hline 
1 & 1  & 1  & 1  & $\begin{array}{c}
62.71\\
-
\end{array}$ & $\begin{array}{c}
417.41\\
-
\end{array}$ & $\begin{array}{c}
1437.3\\
-
\end{array}$ & $\begin{array}{c}
3228.7\\
-
\end{array}$ & $\begin{array}{c}
5218.1\\
-
\end{array}$ & $\begin{array}{c}
6285.7\\
-
\end{array}$ & $\begin{array}{c}
5717.3\\
-
\end{array}$ & $\begin{array}{c}
3879.2\\
-
\end{array}$ & $\begin{array}{c}
1894.7\\
-
\end{array}$ & $\begin{array}{c}
607.32\\
-
\end{array}$ & $\begin{array}{c}
101.69\\
-
\end{array}$\tabularnewline
\hline 
1 & 1  & -1  & -1  & $\begin{array}{c}
88.678\\
-
\end{array}$ & $\begin{array}{c}
427.56\\
-
\end{array}$ & $\begin{array}{c}
1298\\
-
\end{array}$ & $\begin{array}{c}
2637.8\\
-
\end{array}$ & $\begin{array}{c}
4045.7\\
-
\end{array}$ & $\begin{array}{c}
4716.2\\
-
\end{array}$ & $\begin{array}{c}
4307.9\\
-
\end{array}$ & $\begin{array}{c}
2996.2\\
-
\end{array}$ & $\begin{array}{c}
1573.5\\
-
\end{array}$ & $\begin{array}{c}
555.82\\
-
\end{array}$ & $\begin{array}{c}
118.2\\
-
\end{array}$\tabularnewline
\hline \hline
\multicolumn{4}{|c}{Strategy a: } & \multicolumn{11}{c|}{Two Pietarinen expansions }\tabularnewline
\hline 
\hline 
$C_{1}$ & $C_{2}$ & $B_{1}$ & $B_{2}$ & $\begin{array}{c}
c_{0}\\
d_{0}
\end{array}$ & $\begin{array}{c}
c_{1}\\
d_{1}
\end{array}$  & $\begin{array}{c}
c_{2}\\
d_{2}
\end{array}$  & $\begin{array}{c}
c_{3}\\
d_{3}
\end{array}$ & $\begin{array}{c}
c_{4}\\
d_{4}
\end{array}$  & $\begin{array}{c}
c_{5}\\
d_{5}
\end{array}$  & $\begin{array}{c}
c_{6}\\
d_{6}
\end{array}$  & $\begin{array}{c}
c_{7}\\
d_{7}
\end{array}$ & $\begin{array}{c}
c_{8}\\
d_{8}
\end{array}$  & $\begin{array}{c}
c_{9}\\
d_{9}
\end{array}$ & $\begin{array}{c}
c_{10}\\
d_{10}
\end{array}$\tabularnewline
\hline \hline
1  & 0  & 1  & 1  & $\begin{array}{c}
3.849\\
3.849
\end{array}$ & $\begin{array}{c}
5.812\\
-4.385
\end{array}$ & $\begin{array}{c}
-3.141\\
4.217
\end{array}$ & $\begin{array}{c}
-10.608\\
-3.331
\end{array}$ & $\begin{array}{c}
-6.01\\
1.494
\end{array}$ & $\begin{array}{c}
-0.704\\
-0.416
\end{array}$ & $\begin{array}{c}
-\\
-
\end{array}$ & $\begin{array}{c}
-\\
-
\end{array}$ & $\begin{array}{c}
-\\
-
\end{array}$ & $\begin{array}{c}
-\\
-
\end{array}$ & $\begin{array}{c}
-\\
-
\end{array}$\tabularnewline
\hline 
1 & 0  & -1  & -1  & $\begin{array}{c}
1.731\\
1.731
\end{array}$ & $\begin{array}{c}
-2.290\\
-1.212
\end{array}$ & $\begin{array}{c}
-0.002\\
0.573
\end{array}$ & $\begin{array}{c}
-5.747\\
-0.402
\end{array}$ & $\begin{array}{c}
0.693\\
0.194
\end{array}$ & $\begin{array}{c}
-2.229\\
-0.070
\end{array}$ & $\begin{array}{c}
1.239\\
0.018
\end{array}$ & $\begin{array}{c}
-\\
-
\end{array}$ & $\begin{array}{c}
-\\
-
\end{array}$ & $\begin{array}{c}
-\\
-
\end{array}$ & $\begin{array}{c}
-\\
-
\end{array}$\tabularnewline
\hline 
0 & 1  & 1  & 1  & $\begin{array}{c}
18.667\\
18.667
\end{array}$ & $\begin{array}{c}
117.70\\
-8.027
\end{array}$ & $\begin{array}{c}
189.14\\
8.294
\end{array}$ & $\begin{array}{c}
171.86\\
-5.408
\end{array}$ & $\begin{array}{c}
89.342\\
2.066
\end{array}$ & $\begin{array}{c}
22.696\\
-0.441
\end{array}$ & $\begin{array}{c}
-\\
-
\end{array}$ & $\begin{array}{c}
-\\
-
\end{array}$ & $\begin{array}{c}
-\\
-
\end{array}$ & $\begin{array}{c}
-\\
-
\end{array}$ & $\begin{array}{c}
-\\
-
\end{array}$\tabularnewline
\hline 
0 & 1  & -1  & -1  & $\begin{array}{c}
-32.588\\
-32.588
\end{array}$ & $\begin{array}{c}
-217.23\\
-0.676
\end{array}$ & $\begin{array}{c}
-220.56\\
0.347
\end{array}$ & $\begin{array}{c}
30.972\\
-0.195
\end{array}$ & $\begin{array}{c}
174.4\\
0.109
\end{array}$ & $\begin{array}{c}
-5.098\\
-0.055
\end{array}$ & $\begin{array}{c}
-190.34\\
0.025
\end{array}$ & $\begin{array}{c}
-147.91\\
-0.008
\end{array}$ & $\begin{array}{c}
-37.90\\
0.002
\end{array}$ & $\begin{array}{c}
-\\
-
\end{array}$ & $\begin{array}{c}
-\\
-
\end{array}$\tabularnewline
\hline 
 1 & 1 & 1 & 1  & $\begin{array}{c}
-1.255\\
-1.255
\end{array}$ & $\begin{array}{c}
-2.575\\
-0.431
\end{array}$ & $\begin{array}{c}
1.651\\
-0.754
\end{array}$ & $\begin{array}{c}
4.307\\
1.189
\end{array}$ & $\begin{array}{c}
3.093\\
-1.122
\end{array}$ & $\begin{array}{c}
-1.463\\
0.772
\end{array}$ & $\begin{array}{c}
-2.319\\
-0.403
\end{array}$ & $\begin{array}{c}
0.232\\
0.144
\end{array}$ & $\begin{array}{c}
1.907\\
-0.032
\end{array}$ & $\begin{array}{c}
1.221\\
-
\end{array}$ & $\begin{array}{c}
-\\
-
\end{array}$\tabularnewline
\hline 
1 & 1 & -1 & -1 & $\begin{array}{c}
-14.963\\
-14.963
\end{array}$ & $\begin{array}{c}
3.089\\
136.57
\end{array}$ & $\begin{array}{c}
4.009\\
-230.17
\end{array}$ & $\begin{array}{c}
3.351\\
313.67
\end{array}$ & $\begin{array}{c}
3.146\\
-443.08
\end{array}$ & $\begin{array}{c}
1.1584\\
519.08
\end{array}$ & $\begin{array}{c}
0.1584\\
-476.42
\end{array}$ & $\begin{array}{c}
-0.686\\
301.89
\end{array}$ & $\begin{array}{c}
-0.572\\
-87.63
\end{array}$ & $\begin{array}{c}
-0.430\\
-
\end{array}$ & $\begin{array}{c}
-\\
-
\end{array}$\tabularnewline
\hline \hline
\multicolumn{4}{|c}{Strategy a: } & \multicolumn{11}{c|}{Three Pietarinen expansions }\tabularnewline
\hline \hline
$C_{1}$ & $C_{2}$ & $B_{1}$ & $B_{2}$ & $\begin{array}{c}
c_{0}\\
d_{0}\\
e_{0}
\end{array}$ & $\begin{array}{c}
c_{1}\\
d_{1}\\
e_{1}
\end{array}$  & $\begin{array}{c}
c_{2}\\
d_{2}\\
e_{2}
\end{array}$  & $\begin{array}{c}
c_{3}\\
d_{3}\\
e_{3}
\end{array}$ & $\begin{array}{c}
c_{4}\\
d_{4}\\
e_{4}
\end{array}$  & $\begin{array}{c}
c_{5}\\
d_{5}\\
e_{5}
\end{array}$  & $\begin{array}{c}
c_{6}\\
d_{6}\\
e_{6}
\end{array}$  & $\begin{array}{c}
c_{7}\\
d_{7}\\
e_{7}
\end{array}$ & $\begin{array}{c}
c_{8}\\
d_{8}\\
e_{8}
\end{array}$  & $\begin{array}{c}
c_{9}\\
d_{9}\\
e_{9}
\end{array}$ & $\begin{array}{c}
c_{10}\\
d_{10}\\
e_{10}
\end{array}$\tabularnewline
\hline \hline
1 & 1 & 1 & 1 & $\begin{array}{c}
2.275\\
2.274\\
2.277
\end{array}$ & $\begin{array}{c}
19.465\\
-0.616\\
-0.878
\end{array}$  & $\begin{array}{c}
27.86\\
0.044\\
0.063
\end{array}$  & $\begin{array}{c}
35.107\\
-0.023\\
-0.029
\end{array}$ & $\begin{array}{c}
18.749\\
0.019\\
-
\end{array}$  & $\begin{array}{c}
12.32\\
-0.007\\
-
\end{array}$  & $\begin{array}{c}
-\\
-\\
-
\end{array}$  & $\begin{array}{c}
-\\
-\\
-
\end{array}$ & $\begin{array}{c}
-\\
-\\
-
\end{array}$ & $\begin{array}{c}
-\\
-\\
-
\end{array}$ & $\begin{array}{c}
-\\
-\\
-
\end{array}$\tabularnewline
\hline 
1 & 1 & -1 & -1 & $\begin{array}{c}
10.107\\
10.107\\
10.107
\end{array}$ & $\begin{array}{c}
1.372\\
-70.87\\
-0.757
\end{array}$ & $\begin{array}{c}
0.849\\
65.949\\
0.077
\end{array}$ & $\begin{array}{c}
-0.515\\
-21.659\\
-0.017
\end{array}$ & $\begin{array}{c}
-\\
-\\
-
\end{array}$ & $\begin{array}{c}
-\\
-\\
-
\end{array}$ & $\begin{array}{c}
-\\
-\\
-
\end{array}$ & $\begin{array}{c}
-\\
-\\
-
\end{array}$ & $\begin{array}{c}
-\\
-\\
-
\end{array}$ & $\begin{array}{c}
-\\
-\\
-
\end{array}$ & $\begin{array}{c}
-\\
-\\
-
\end{array}$\tabularnewline
\hline \hline
\end{tabular}
\end{table*}

\begin{table*}[t!]
{Pietarinen expansion coefficients corresponding to solutions presented
in Table \ref{tb1:Toy model parameters} \vspace*{0.5cm} (\textbf{Strategy b}).}
\label{tb3:Coefficients Strategy b} 
\footnotesize

\begin{tabular}{|c|c|c|c|ccccccccccc|}
\hline \hline
\multicolumn{4}{|c}{Startegy b:} & \multicolumn{11}{c|}{One Pietarinen expansion}\tabularnewline
\hline \hline
$C_{1}$ & $C_{2}$ & $B_{1}$ & $B_{2}$ & $\begin{array}{c}
c_{0}\\
c_{11}
\end{array}$ & $\begin{array}{c}
c_{1}\\
c_{12}
\end{array}$  & $\begin{array}{c}
c_{2}\\
c_{13}
\end{array}$  & $\begin{array}{c}
c_{3}\\
c_{14}
\end{array}$ & $\begin{array}{c}
c_{4}\\
c_{15}
\end{array}$  & $\begin{array}{c}
c_{5}\\
c_{16}
\end{array}$  & $\begin{array}{c}
c_{6}\\
c_{17}
\end{array}$  & $\begin{array}{c}
c_{7}\\
c_{18}
\end{array}$ & $\begin{array}{c}
c_{8}\\
c_{19}
\end{array}$  & $\begin{array}{c}
c_{9}\\
c_{20}
\end{array}$ & $\begin{array}{c}
c_{10}\\
c_{21}
\end{array}$\tabularnewline
\hline 
1 & 0  & 0  & 0  & $\begin{array}{c}
0.676\\
-2.938
\end{array}$ & $\begin{array}{c}
-0.795\\
2.575
\end{array}$ & $\begin{array}{c}
0.947\\
-1.934
\end{array}$ & $\begin{array}{c}
-1.948\\
1.520
\end{array}$ & $\begin{array}{c}
2.450\\
-0.864
\end{array}$ & $\begin{array}{c}
-3.236\\
0.553
\end{array}$ & $\begin{array}{c}
4.100\\
-0.248
\end{array}$ & $\begin{array}{c}
-4.084\\
0.121
\end{array}$ & $\begin{array}{c}
4.034\\
-
\end{array}$ & $\begin{array}{c}
-3.702\\
-
\end{array}$ & $\begin{array}{c}
3.640\\
-
\end{array}$\tabularnewline
\hline 
0  & 1  & 0  & 0  & $\begin{array}{c}
-0.362\\
-
\end{array}$ & $\begin{array}{c}
0.536\\
-
\end{array}$ & $\begin{array}{c}
-0.410\\
-
\end{array}$ & $\begin{array}{c}
0.211\\
-
\end{array}$ & $\begin{array}{c}
-0.170\\
-
\end{array}$ & $\begin{array}{c}
0.132\\
-
\end{array}$ & $\begin{array}{c}
-0.042\\
-
\end{array}$ & $\begin{array}{c}
0.074\\
-
\end{array}$ & $\begin{array}{c}
-0.011\\
-
\end{array}$ & $\begin{array}{c}
0.021\\
-
\end{array}$ & $\begin{array}{c}
-\\
-
\end{array}$\tabularnewline
\hline 
0  & 0  & 1  & 1  & $\begin{array}{c}
-0.560\\
-
\end{array}$ & $\begin{array}{c}
-1.205\\
-
\end{array}$ & $\begin{array}{c}
-2.397\\
-
\end{array}$ & $\begin{array}{c}
-1.364\\
-
\end{array}$ & $\begin{array}{c}
-0.517\\
-
\end{array}$ & $\begin{array}{c}
-1.170\\
-
\end{array}$ & $\begin{array}{c}
-0.612\\
-
\end{array}$ & $\begin{array}{c}
-0.193\\
-
\end{array}$ & $\begin{array}{c}
-0.199\\
-
\end{array}$ & $\begin{array}{c}
-\\
-
\end{array}$ & $\begin{array}{c}
-\\
-
\end{array}$\tabularnewline
\hline 
0  & 0  & -1  & -1  & $\begin{array}{c}
0.564\\
-
\end{array}$ & $\begin{array}{c}
-0.020\\
-
\end{array}$ & $\begin{array}{c}
1.994\\
-
\end{array}$ & $\begin{array}{c}
1.060\\
-
\end{array}$ & $\begin{array}{c}
0.168\\
-
\end{array}$ & $\begin{array}{c}
2.039\\
-
\end{array}$ & $\begin{array}{c}
0.397\\
-
\end{array}$ & $\begin{array}{c}
0.598\\
-
\end{array}$ & $\begin{array}{c}
1.357\\
-
\end{array}$ & $\begin{array}{c}
-\\
-
\end{array}$ & $\begin{array}{c}
-\\
-
\end{array}$\tabularnewline
\hline 
1 & 0  & 1  & 1  & $\begin{array}{c}
1.255\\
0.245
\end{array}$ & $\begin{array}{c}
1.799\\
-
\end{array}$ & $\begin{array}{c}
2.300\\
-
\end{array}$ & $\begin{array}{c}
1.412\\
-
\end{array}$ & $\begin{array}{c}
1.418\\
-
\end{array}$ & $\begin{array}{c}
2.447\\
-
\end{array}$ & $\begin{array}{c}
1.143\\
-
\end{array}$ & $\begin{array}{c}
1.178\\
-
\end{array}$ & $\begin{array}{c}
1.134\\
-
\end{array}$ & $\begin{array}{c}
0.484\\
-
\end{array}$ & $\begin{array}{c}
0.300\\
-
\end{array}$\tabularnewline
\hline 
1  & 0  & - 1  & - 1  & $\begin{array}{c}
0.777\\
1.029
\end{array}$ & $\begin{array}{c}
0.986\\
0.987
\end{array}$ & $\begin{array}{c}
1.832\\
0.782
\end{array}$ & $\begin{array}{c}
1.284\\
1.281
\end{array}$ & $\begin{array}{c}
1.641\\
1.237
\end{array}$ & $\begin{array}{c}
0.661\\
0.316
\end{array}$ & $\begin{array}{c}
0.807\\
1.570
\end{array}$ & $\begin{array}{c}
1.056\\
1.216
\end{array}$ & $\begin{array}{c}
0.709\\
-0.781
\end{array}$ & $\begin{array}{c}
1.583\\
-0.934
\end{array}$ & $\begin{array}{c}
0.617\\
-
\end{array}$\tabularnewline
\hline 
0 & 1  & 1  & 1  & $\begin{array}{c}
1.388\\
-
\end{array}$ & $\begin{array}{c}
1.753\\
-
\end{array}$ & $\begin{array}{c}
5.598\\
-
\end{array}$ & $\begin{array}{c}
6.395\\
-
\end{array}$ & $\begin{array}{c}
11.384\\
-
\end{array}$ & $\begin{array}{c}
10.090\\
-
\end{array}$ & $\begin{array}{c}
13.046\\
-
\end{array}$ & $\begin{array}{c}
9.156\\
-
\end{array}$ & $\begin{array}{c}
9.852\\
-
\end{array}$ & $\begin{array}{c}
3.338\\
-
\end{array}$ & $\begin{array}{c}
2.805\\
-
\end{array}$\tabularnewline
\hline 
0  & 1  & -1  & -1  & $\begin{array}{c}
1.565\\
-
\end{array}$ & $\begin{array}{c}
-0.031\\
-
\end{array}$ & $\begin{array}{c}
-0.235\\
-
\end{array}$ & $\begin{array}{c}
1.319\\
-
\end{array}$ & $\begin{array}{c}
1.635\\
-
\end{array}$ & $\begin{array}{c}
1.836\\
-
\end{array}$ & $\begin{array}{c}
1.316\\
-
\end{array}$ & $\begin{array}{c}
1.179\\
-
\end{array}$ & $\begin{array}{c}
0.392\\
-
\end{array}$ & $\begin{array}{c}
0.381\\
-
\end{array}$ & $\begin{array}{c}
-\\
-
\end{array}$\tabularnewline
\hline 
1 & 1  & 1  & 1  & $\begin{array}{c}
-0.088\\
1.242
\end{array}$ & $\begin{array}{c}
1.193\\
0.037
\end{array}$ & $\begin{array}{c}
-0.502\\
-
\end{array}$ & $\begin{array}{c}
2.507\\
-
\end{array}$ & $\begin{array}{c}
-0.769\\
-
\end{array}$ & $\begin{array}{c}
3.256\\
-
\end{array}$ & $\begin{array}{c}
-1.326\\
-
\end{array}$ & $\begin{array}{c}
5.019\\
-
\end{array}$ & $\begin{array}{c}
-0.963\\
-
\end{array}$ & $\begin{array}{c}
2.993\\
-
\end{array}$ & $\begin{array}{c}
-0.645\\
-
\end{array}$\tabularnewline
\hline 
1 & 1  & -1  & -1  & $\begin{array}{c}
4.426\\
2.423
\end{array}$ & $\begin{array}{c}
-1.681\\
-1.484
\end{array}$ & $\begin{array}{c}
3.943\\
-
\end{array}$ & $\begin{array}{c}
-3.801\\
-
\end{array}$ & $\begin{array}{c}
3.305\\
-
\end{array}$ & $\begin{array}{c}
0.263\\
-
\end{array}$ & $\begin{array}{c}
-0.650\\
-
\end{array}$ & $\begin{array}{c}
6.418\\
-
\end{array}$ & $\begin{array}{c}
-4.716\\
-
\end{array}$ & $\begin{array}{c}
6.584\\
-
\end{array}$ & $\begin{array}{c}
-4.286\\
-
\end{array}$\tabularnewline
\hline \hline
\multicolumn{4}{|c}{Strategy b: } & \multicolumn{11}{c|}{Two Pietarinen expansions }\tabularnewline
\hline 
\hline 
$C_{1}$ & $C_{2}$ & $B_{1}$ & $B_{2}$ & $\begin{array}{c}
c_{0}\\
d_{0}
\end{array}$ & $\begin{array}{c}
c_{1}\\
d_{1}
\end{array}$  & $\begin{array}{c}
c_{2}\\
d_{2}
\end{array}$  & $\begin{array}{c}
c_{3}\\
d_{3}
\end{array}$ & $\begin{array}{c}
c_{4}\\
d_{4}
\end{array}$  & $\begin{array}{c}
c_{5}\\
d_{5}
\end{array}$  & $\begin{array}{c}
c_{6}\\
d_{6}
\end{array}$  & $\begin{array}{c}
c_{7}\\
d_{7}
\end{array}$ & $\begin{array}{c}
c_{8}\\
d_{8}
\end{array}$  & $\begin{array}{c}
c_{9}\\
d_{9}
\end{array}$ & $\begin{array}{c}
c_{10}\\
d_{10}
\end{array}$\tabularnewline
\hline \hline
1  & 0  & 1  & 1  & $\begin{array}{c}
0.512\\
0.512
\end{array}$ & $\begin{array}{c}
1.761\\
-0.467
\end{array}$ & $\begin{array}{c}
1.072\\
0.104
\end{array}$ & $\begin{array}{c}
-0.373\\
-0.375
\end{array}$ & $\begin{array}{c}
1.552\\
0.148
\end{array}$ & $\begin{array}{c}
1.780\\
-0.039
\end{array}$ & $\begin{array}{c}
-\\
-
\end{array}$ & $\begin{array}{c}
-\\
-
\end{array}$ & $\begin{array}{c}
-\\
-
\end{array}$ & $\begin{array}{c}
-\\
-
\end{array}$ & $\begin{array}{c}
-\\
-
\end{array}$\tabularnewline
\hline 
1 & 0  & -1  & -1  & $\begin{array}{c}
0.301\\
0.301
\end{array}$ & $\begin{array}{c}
-2.718\\
0.455
\end{array}$ & $\begin{array}{c}
3.135\\
-2.287
\end{array}$ & $\begin{array}{c}
0.373\\
-1.714
\end{array}$ & $\begin{array}{c}
-3.393\\
4.815
\end{array}$ & $\begin{array}{c}
1.769\\
-3.150
\end{array}$ & $\begin{array}{c}
2.913\\
0.837
\end{array}$ & $\begin{array}{c}
-\\
-
\end{array}$ & $\begin{array}{c}
-\\
-
\end{array}$ & $\begin{array}{c}
-\\
-
\end{array}$ & $\begin{array}{c}
-\\
-
\end{array}$\tabularnewline
\hline 
0 & 1  & 1  & 1  & $\begin{array}{c}
0.793\\
0.793
\end{array}$ & $\begin{array}{c}
0.662\\
0.572
\end{array}$ & $\begin{array}{c}
0.629\\
-0.120
\end{array}$ & $\begin{array}{c}
1.267\\
0.035
\end{array}$ & $\begin{array}{c}
1.042\\
-0.011
\end{array}$ & $\begin{array}{c}
0.081\\
0.003
\end{array}$ & $\begin{array}{c}
-\\
-
\end{array}$ & $\begin{array}{c}
-\\
-
\end{array}$ & $\begin{array}{c}
-\\
-
\end{array}$ & $\begin{array}{c}
-\\
-
\end{array}$ & $\begin{array}{c}
-\\
-
\end{array}$\tabularnewline
\hline 
0 & 1  & -1  & -1  & $\begin{array}{c}
-1.649\\
-1.649
\end{array}$ & $\begin{array}{c}
-12.558\\
--29.437
\end{array}$ & $\begin{array}{c}
2.551\\
8.936
\end{array}$ & $\begin{array}{c}
20.807\\
12.801
\end{array}$ & $\begin{array}{c}
-4.393\\
12.373
\end{array}$ & $\begin{array}{c}
-17.838\\
-25.197
\end{array}$ & $\begin{array}{c}
-14.688\\
7.424
\end{array}$ & $\begin{array}{c}
-\\
-
\end{array}$ & $\begin{array}{c}
-\\
-
\end{array}$ & $\begin{array}{c}
-\\
-
\end{array}$ & $\begin{array}{c}
-\\
-
\end{array}$\tabularnewline
\hline 
 1 & 1 & 1 & 1  & $\begin{array}{c}
6.410\\
6.410
\end{array}$ & $\begin{array}{c}
33.756\\
1.223
\end{array}$ & $\begin{array}{c}
55.745\\
-0.903
\end{array}$ & $\begin{array}{c}
59.202\\
0.437
\end{array}$ & $\begin{array}{c}
44.944\\
-0.194
\end{array}$ & $\begin{array}{c}
21.583\\
0.060
\end{array}$ & $\begin{array}{c}
6.587\\
-0.171
\end{array}$ & $\begin{array}{c}
-\\
-
\end{array}$ & $\begin{array}{c}
-\\
-
\end{array}$ & $\begin{array}{c}
-\\
-
\end{array}$ & $\begin{array}{c}
-\\
-
\end{array}$\tabularnewline
\hline 
1 & 1 & -1 & -1 & $\begin{array}{c}
1.960\\
1.960
\end{array}$ & $\begin{array}{c}
-2.298\\
-0.817
\end{array}$ & $\begin{array}{c}
2.642\\
0.279
\end{array}$ & $\begin{array}{c}
-2.579\\
-0.581
\end{array}$ & $\begin{array}{c}
2.374\\
0.392
\end{array}$ & $\begin{array}{c}
-1.659\\
0.115
\end{array}$ & $\begin{array}{c}
1.027\\
-0.197
\end{array}$ & $\begin{array}{c}
-0.424\\
0.068
\end{array}$ & $\begin{array}{c}
0.142\\
-
\end{array}$ & $\begin{array}{c}
-\\
-
\end{array}$ & $\begin{array}{c}
-\\
-
\end{array}$\tabularnewline
\hline \hline
\multicolumn{4}{|c}{Strategy b: } & \multicolumn{11}{c|}{Three Pietarinen expansions }\tabularnewline
\hline \hline
$C_{1}$ & $C_{2}$ & $B_{1}$ & $B_{2}$ & $\begin{array}{c}
c_{0}\\
d_{0}\\
e_{0}
\end{array}$ & $\begin{array}{c}
c_{1}\\
d_{1}\\
e_{1}
\end{array}$  & $\begin{array}{c}
c_{2}\\
d_{2}\\
e_{2}
\end{array}$  & $\begin{array}{c}
c_{3}\\
d_{3}\\
e_{3}
\end{array}$ & $\begin{array}{c}
c_{4}\\
d_{4}\\
e_{4}
\end{array}$  & $\begin{array}{c}
c_{5}\\
d_{5}\\
e_{5}
\end{array}$  & $\begin{array}{c}
c_{6}\\
d_{6}\\
e_{6}
\end{array}$  & $\begin{array}{c}
c_{7}\\
d_{7}\\
e_{7}
\end{array}$ & $\begin{array}{c}
c_{8}\\
d_{8}\\
e_{8}
\end{array}$  & $\begin{array}{c}
c_{9}\\
d_{9}\\
e_{9}
\end{array}$ & $\begin{array}{c}
c_{10}\\
d_{10}\\
e_{10}
\end{array}$\tabularnewline
\hline \hline
1 & 1 & 1 & 1 & $\begin{array}{c}
-1.608\\
-1.608\\
-1.608
\end{array}$ & $\begin{array}{c}
-5.203\\
0.515\\
1.518
\end{array}$ & $\begin{array}{c}
5.819\\
0.252\\
1.380
\end{array}$ & $\begin{array}{c}
1.520\\
-1.565\\
-0.229
\end{array}$ & $\begin{array}{c}
-3.035\\
-2.579\\
-0.321
\end{array}$ & $\begin{array}{c}
4.982\\
-2.229\\
0.411
\end{array}$ & $\begin{array}{c}
5.711\\
-1.062\\
0.598
\end{array}$ & $\begin{array}{c}
-0.305\\
-0.267\\
0.282
\end{array}$ & $\begin{array}{c}
-\\
-
\end{array}$ & $\begin{array}{c}
-\\
-
\end{array}$ & $\begin{array}{c}
-\\
-
\end{array}$\tabularnewline
\hline 
1 & 1 & -1 & -1 & $\begin{array}{c}
1.842\\
1.842\\
1.842
\end{array}$ & $\begin{array}{c}
19.214\\
11.827\\
-5.396
\end{array}$ & $\begin{array}{c}
40.571\\
24.661\\
17.925
\end{array}$ & $\begin{array}{c}
49.783\\
-54.662\\
-20.529
\end{array}$ & $\begin{array}{c}
42.630\\
49.917\\
11.084
\end{array}$ & $\begin{array}{c}
32.165\\
-30.450\\
-2.477
\end{array}$ & $\begin{array}{c}
6.188\\
12.023\\
-
\end{array}$ & $\begin{array}{c}
3.091\\
-2.268\\
-
\end{array}$ & $\begin{array}{c}
-\\
-
\end{array}$ & $\begin{array}{c}
-\\
-
\end{array}$ & $\begin{array}{c}
-\\
-
\end{array}$\tabularnewline
\hline \hline
\end{tabular}
\end{table*}

\end{document}